\documentclass[aps,twocolumn,showpacs,preprintnumbers,prb,superscriptaddress]{revtex4}

\usepackage{graphicx}
\usepackage{amssymb}

\newcommand{\R}{\mathbf{r}}

\newcommand{\UP}{n_{\uparrow}}
\newcommand{\DN}{n_{\downarrow}}

\newcommand{\be}{\begin{equation}}
\newcommand{\ee}{\end{equation}}
\newcommand{\bea}{\begin{eqnarray}}
\newcommand{\eea}{\end{eqnarray}}
\newcommand{\bean}{\begin{eqnarray*}}
\newcommand{\eean}{\end{eqnarray*}}

\begin{document}

\title{Semiclassical atom theory applied to solid-state physics}

\author{Lucian A. Constantin}
\affiliation{Center for Biomolecular Nanotechnologies @UNILE, Istituto Italiano di Tecnologia, Via Barsanti, I-73010 Arnesano, Italy}
\author{Aleksandrs Terentjevs}
\affiliation{Istituto Nanoscienze-CNR, Euromediterranean Center for Nanomaterial Modelling and Technology (ECMT), Via per Arnesano 73100 Lecce, Italy }
\author{Fabio Della Sala}
\affiliation{Istituto Nanoscienze-CNR, Euromediterranean Center for Nanomaterial Modelling and Technology (ECMT), Via per Arnesano 73100 Lecce, Italy }
\affiliation{Center for Biomolecular Nanotechnologies @UNILE, Istituto Italiano di Tecnologia, Via Barsanti, I-73010 Arnesano, Italy}
\author{Pietro Cortona}
\affiliation{Laboratoire Structures, Propri\'et\'es et Mod\'elisation des Solides, CNRS UMR 8580, \'Ecole Centrale Paris, Grand Voie des Vignes, F-92295 Ch\^{a}tenay-Malabry, France}
\author{Eduardo Fabiano}
\affiliation{Istituto Nanoscienze-CNR, Euromediterranean Center for Nanomaterial Modelling and Technology (ECMT), Via per Arnesano 73100 Lecce, Italy }
\affiliation{Center for Biomolecular Nanotechnologies @UNILE, Istituto Italiano di Tecnologia, Via Barsanti, I-73010 Arnesano, Italy}

\date{\today}

\begin{abstract}
Using the semiclassical neutral atom theory,
we extend to fourth order the modified gradient expansion 
of the exchange energy of density functional theory.
This expansion can be applied both to large atoms
and solid-state problems. Moreover, we show that it
can be employed to construct a simple and non-empirical 
generalized gradient approximation (GGA) exchange-correlation functional
competitive with state-of-the-art GGAs for solids, but also
reasonably accurate for large atoms and ordinary chemistry. 
\end{abstract}

\pacs{71.10.Ca,71.15.Mb,71.45.Gm}

\maketitle

\section{Introduction}

Density functional theory (DFT) \cite{hk,KS,dobson_vignale_book}
is one of the most popular computational approaches
to material science and condensed-matter physics. 
However, the final accuracy of DFT calculations depends on the
approximation used for the exchange-correlation (XC)
functional, which describes the quantum effects on the
electron-electron interaction.
Thus, the development and testing of new XC functionals have
been  active research fields during the last decades 
\cite{scusrev,rappoport09,civalleri}.

Model systems are fundamental tools for the development
of non-empirical DFT functionals.
One popular model is the electron gas with slowly-varying density.
Performing a second-order gradient expansion (GE2) of the
exchange energy density $\epsilon_x$, this model gives
\begin{equation}
\epsilon_x = \epsilon_x^{LDA}(1+\mu^{GE2}s^2)\ ,
\end{equation}
where $\epsilon_x^{LDA}=-(3/4)(3/\pi)^{1/3}n^{4/3}$ 
is the exchange energy density in the local density 
approximation (LDA) \cite{KS},
$n$ is the electron density, $s=|\nabla n|/[2(3\pi^2)^{1/3}n^{4/3}]$ is the 
dimensionless reduced gradient, and 
$\mu^{GE2}=10/81$ is the GE2 coefficient. 
The slowly varying density regime is considered a paradigm
for solid-state physics, and the GE2 has been successfully used
as a key tool to develop 
Generalized Gradient Approximation (GGA)
functionals
\cite{PBEsol,sogga,sogga11,PBEint,vt84,vela09,pedroza09,RGE2,inttca}
as well as  meta-GGA functionals 
\cite{TPSS,revTPSS,MGGAMS2,metavt84,BLOC,SCAN,M11L}.

Another important model system is the semiclassical neutral atom (SCA),
whose theory was established several years ago 
\cite{thomas26,fermi27,scott52,schwinger80,schwinger81,englert84,englert85}. 
This model has been recently used to
derive a modified second-order gradient expansion (MGE2) 
for exchange \cite{EB09}
\begin{equation}\label{eee2}
\epsilon_x = \epsilon_x^{LDA}(1+\mu^{MGE2}s^2)\ ,
\end{equation}
where $\mu^{MGE2}=0.26$. This expansion has been shown to be relevant for
the accurate DFT description of atoms and molecules
\cite{EB09,ELCB,lee09,PCSB,APBE,hapbe,ioni}.

The two gradient expansions discussed above
have been employed to
develop GGA functionals accurate either for solid-state
(e.g. the PBEsol functional of Ref. \onlinecite{PBEsol}),
or for chemistry (e.g. the APBE of Ref. \onlinecite{APBE}).
Nevertheless, the exchange enhancement factor $F_x$ (defined by 
$E_x[n]=\int d\R\;\epsilon_x^{LDA}F_x(s)$) of both
both PBEsol and APBE behaves as
\begin{equation}
F_x(s\rightarrow 0)\rightarrow 1+\mu s^2+\nu s^4+\mathcal{O}\left(s^6\right) \; \textrm{with}\;  \nu=-\frac{\mu^2}{\kappa}\ ,
\label{e1}
\end{equation}
where $\kappa=0.804$ is
fixed from the Lieb-Oxford bound \cite{lieb81,sLL},
and $\mu$ is the pertinent second-order coefficient.
The importance of the fourth-order term in this development 
has been already discussed in literature \cite{madsen,RGE2,wc,wcomm}.
It is important for intermediate values of the reduced gradient
($0.3\lessapprox s\lessapprox 1$) as those often
encountered in bulk solids.
Indeed, functionals which recover GE2, but with the $s^4$ 
term in the Taylor expansion of exchange set to zero, e.g. 
the ones in Refs \onlinecite{RGE2,PBEint}, show a quite different behavior 
with respect to PBEsol (see also results in Ref. \onlinecite{inttca}).
On the other hand,  the relevance of higher-order terms
in the modified gradient expansion has not yet
been investigated.

In this work we consider this issue and 
we use the SCA theory to introduce an extension of the modified
gradient expansion to fourth-order (MGE4). 
We show that this expansion is appropriate for both 
semiclassical atoms and solid-state problems.
Moreover, a simple GGA, based on MGE4, is constructed.
These achievements will emphasize the importance of the SCA model for DFT,
not only when one is concerned with finite systems \cite{APBE}, but also
in the popular field of solid-state physics.

\section{Theory}
In Ref. \onlinecite{EB09}, the MGE2 coefficient
was derived by requiring that the expansion given in Eq. (\ref{eee2})
has to be large-Z asymptotically exact to the 1st degree.
Thus, in practice, an energy constraint was applied and
the MGE2 was forced to recover the correct lowest
coefficient of the semiclassical expansion
of the exchange energy. 

An alternative way to derive the MGE2, is to impose
that, in the slowly-varying density region of non-relativistic large neutral
atoms (i.e. with $Z\rightarrow\infty$),
the modified gradient expansion recovers the exact exchange energy.
Thus, for $Z\rightarrow\infty$, we have
\begin{eqnarray}
\int_V\epsilon_x^{exact}d\R = \int_V\epsilon_x^{LDA} (1+\mu s^2+ \nu s^4+\ldots)d\R .
\label{eee1}
\end{eqnarray}
The integration is performed on the
slowly-varying density region $V$, defined by the condition
$-1\leq q\leq 1$,
where  $q=\nabla^2 n/\{4(3\pi^{2})^{2/3}n^{5/3}\}$.
Note that this region dominates for an atom
with an infinite number of electrons and it is also the only
one where a gradient expansion makes sense.
The use of the reduced Laplacian $q$ in the definition
of $V$ is motivated by the fact that
slowly-varying density regions
of atoms cannot be defined in terms
of $s$ only, because the reduced gradient is small also
near the nuclear cusp \cite{alpha}, where the density is rapidly varying.
Instead, they are well identified by considering the
reduced Laplacian $q$
and using the condition $|q| \approx s^2 \lessapprox 1$
(conversely $q\rightarrow -\infty$ near the nucleus and
$q\rightarrow \infty$ in the tail).

The semiclassical theory of atoms is based on the Thomas-Fermi scaling \cite{ELCB},
which implies the following scaling rules for the density and 
the reduced gradients:
\begin{equation}\label{e35}
\begin{array}{lcl}
n_\lambda(\R) = \lambda^2n(\lambda^{1/3}\R)  &\; , \;&  r_{s\lambda}(\R) = \lambda^{-2/3}r_s(\lambda^{1/3}\R)\ ,\\
s_\lambda(\R) = \lambda^{-1/3}s(\lambda^{1/3}\R) &\; , \;& q_\lambda(\R) = \lambda^{-2/3}q(\lambda^{1/3}\R)\ ,
\end{array}
\end{equation}
with $\lambda\rightarrow\infty$, while the nuclear
charge behaves as $Z\rightarrow \lambda Z$, in order to preserve
the total charge neutrality. 
In Eq. (\ref{e35}), $r_s=(3/4\pi n)^{1/3}$ is the Wigner-Seitz radius.
Using these scaling relations, we obtain (in the limit $\lambda\rightarrow\infty$)
\begin{equation}\label{emge2}
\mu^{MGE2}=\lim_{Z\rightarrow\infty}
\frac{\int_V(\epsilon_x^{exact}[Z]-\epsilon_x^{LDA}[Z])d\R}{\int_V\epsilon_x^{LDA}[Z]s[Z]^2d\R}\ .
\end{equation}
Here $[Z]$ denotes that all quantities are computed for the non-relativistic
atom with $Z$ electrons. 

\begin{figure}
\includegraphics[width=\columnwidth]{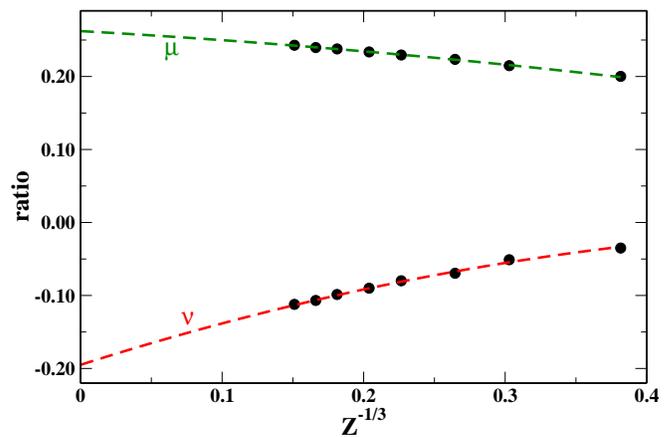}
\caption{Right-hand-side ratios of Eqs. (\ref{emge2}) and (\ref{e2}) (i.e. $\mu$ and $\nu$ values) computed for different noble-gas atoms up to $Z=290$. The extrapolation to $Z\rightarrow\infty$ was done using a parabolic fit as in Refs. \onlinecite{EB09,ioni,PCSB}.}
\label{f1}
\end{figure}
The values of the ratio on the right hand side of Eq. (\ref{emge2})
for different noble-gas atoms, up to $Z=290$, 
are reported in Fig. \ref{f1} together with the 
extrapolation to $Z\rightarrow\infty$. 
We obtain $\mu^{MGE2}=0.262$, that is practically the same 
result as in Ref. \onlinecite{EB09}.

In principle, Eq. (\ref{eee1}) could be also used to obtain 
higher-order results. Nevertheless, such 
an approach is prone to large oscillations
because the integral of $\epsilon_x^{exact}-\epsilon_x^{LDA}(1+\mu^{MGE2}s^2)$
is small by construction, as it is that of $\epsilon_x^{LDA}s^4$, 
for $Z\rightarrow\infty$.
Thus, in order to extend the modified gradient expansion
to fourth order, we need to impose an additional constraint
beyond the energy one.
We do this by requiring that the modified gradient expansion
not only reproduces the SCA asymptotic energy (which yields MGE2),
but also gives a realistic SCA enhancement factor 
in the slowly-varying density limit.
With this choice we also obtain to reduce the importance of high-density
regions, which instead dominate the MGE2 behavior, since they are the ones that
mostly contribute to the energy density, Thus, we can achieve a more
balanced description of the whole slowly-varying regime (including low-density
regions, which are rather important in real bulk solids).
We recall that the enhancement factor is not an
observable and is defined only up to a gauge transformation \cite{pb}.
Nevertheless, the enhancement factor 
$F_x^{exact}=\epsilon_x^{exact}/\epsilon_x^{LDA}$
of the conventional exact exchange
energy density is well defined and has a clear physical meaning  (i.e.
it measures the interaction
between an electron and the true exchange hole).
Note also that the non-uniqueness problem is reduced in the slowly-varying
density limit, where the exact exchange hole has a
semilocal expansion \cite{Becke,ernzerhof98,hole} which becomes unique for the uniform electron gas. Thus, $F_x^{exact}$
can be safely used as a reference for our scope.

Following Eq. (\ref{emge2}), we define
an effective fourth-order coefficient (in the spirit of Eq. (\ref{e1})) as
\begin{equation}
\nu^{MGE4}=\lim_{Z\rightarrow\infty}\frac{\int_V d\R\;(F_x^{exact}[Z]-F_x^{MGE2}[Z])}{\int_V d\R \; s[Z]^4}\ .
\label{e2}
\end{equation}
We remark that any fourth-order gradient expansion of the 
exchange energy diverges for
atoms, because of the exponential decay of the density.
Using the integration technique proposed in
Eq. (\ref{e2}), we remove this difficulty and we focus on the slowly-varying density
regime, that is the only one where a gradient expansion is well defined.

Finally, we highlight that the fourth-order gradient
expansion of exchange depends, in general, on both the
reduced gradient and the Laplacian. However, a Laplacian
dependence is beyond the scope of this paper. Instead,
using Eq. (\ref{e2}), we aim to extract an effective fourth-order
coefficient in the spirit of Eq. (\ref{e1}).

The values of the ratio on the right hand side of Eq. (\ref{e2})
for different noble-gas atoms, up to $Z=290$, 
are reported in Fig. \ref{f1}
together with the extrapolation to $Z\rightarrow\infty$.
The behavior with $Z$ is regular, showing the physical
meaningfulness of Eq. (\ref{e2}). Extrapolation
to $Z\rightarrow\infty$ gives
\begin{equation}
\nu^{MGE4}=-0.195.
\label{e3}
\end{equation}
Note that this limit value is independent on the exact
values used to define the boundaries of region $V$.
Indeed, the same $\nu^{MGE4}$ is obtained
using $-0.8\leq q\leq 0.8$ or $-1.2\leq q\leq 1.2$ 
(plots not reported).

The coefficient of Eq. (\ref{e3}), together with
$\mu^{MGE2}$, define the modified fourth-order gradient
expansion (MGE4), with the following enhancement factor
\begin{equation}
F_x^{MGE4} = 1 + \mu^{MGE2}s^2 + \nu^{MGE4}s^4\ .
\end{equation}
This reproduces, as close as possible, 
the conventional exact exchange enhancement
factor in the slowly-varying density regime
of non-relativistic large neutral atoms.
Note that $\nu^{MGE4}$ is rather different from the fourth-order
coefficient that is implicitly employed in APBE
($-(\mu^{MGE2})^2/\kappa=-0.084$).

\begin{figure}
\includegraphics[width=\columnwidth]{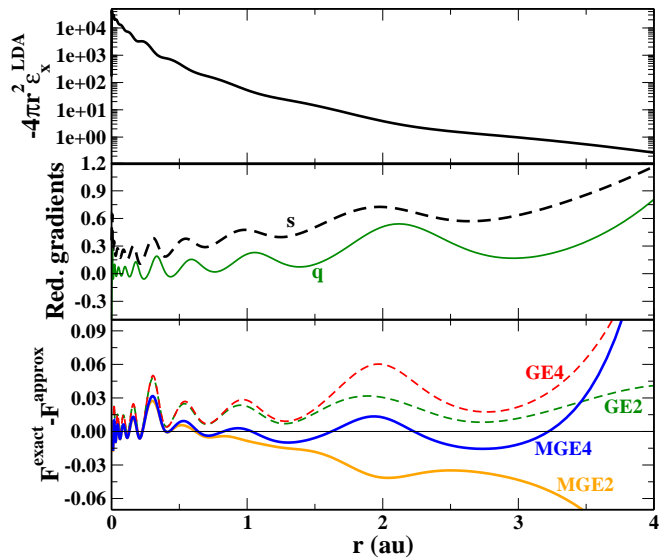}
\caption{\label{fig290} Radial LDA exchange energy density (top panel), reduced gradients $s$ and $q$ (middle panel), and (lower panel) the deviation, with respect to the exact conventional one, of the exchange enhancement factors of several gradient expansions, as functions of the distance $r$ from the nucleus, for the non-relativistic noble atom with 290 electrons.}
\end{figure}

The main features of the MGE4 can be seen in Fig.
\ref{fig290} where we plot, 
for the non-relativistic
noble atom with 290 electrons and in the region $V$ defined
before: the radial LDA exchange energy density,
the reduced gradients $s$ and $q$, and
the deviation, with respect to the
exact conventional one, of the exchange 
enhancement factors of several gradient expansions.
Here GE4 is the conventional fourth-order gradient expansion defined by the
enhancement factor $F_x^{GE4}=1+10/81 s^2+146/2025 q^2-73/405 q s^2 +D s^4$,
where $D=0$ is the best numerical estimation for this parameter
\cite{TPSS}.
We observe that:

$i)$ In the inner atomic core (i.e. for  $r \lessapprox 1$), MGE2 is very accurate
and, because both $s$ and $q$ are relatively small, 
the forth-order terms in gradient expansions are not much significant.
Consequently, MGE4 is as accurate as MGE2.
On the other hand, the GE2 and GE4 exchange enhancement factors are 
smaller, on average, than the exact one and they 
do not describe with high accuracy this energetical region.
In fact, despite both $s$ and $q$ are not large in this
region, only $q$ is very close to zero, while $s\approx 0.3$.
Therefore, the conventional gradient expansions do not work
very well in this regime.
Note that the inner-core high-density region gives the main contribution 
to the exchange energy (99.3\% of it).

$ii)$ In the outer atomic core (i.e. for $1\lessapprox r\lessapprox 4$),
the reduced gradients $s$ and $q$ start to increase, but they are 
still smaller than 1, thus the density is still slowly-varying. 
This atomic region is not very important for the total exchange energy 
of atoms, but it is a model for solid-state problems 
(where the high-density limit is not common). 
While MGE2 fails in this region, the MGE4 and GE2 
are very accurate.

\subsection{Assessment of MGE4 for jellium cluster models}
The MGE4 is accurate by construction for the 
slowly-varying density regime of large non-relativistic
neutral atoms. To test it on a different model, we consider
its performance for jellium clusters. These systems satisfy
the uniform-electron-gas scaling \cite{scaling} 
($n_\lambda(\R)=n(\lambda^{-1/3}\R)$) and, in the limit
of a large number of electrons, they are representative
for solid-state systems.

\begin{figure}
\includegraphics[width=\columnwidth]{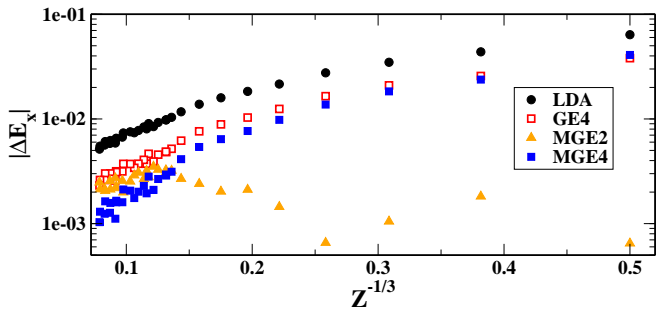}
\caption{Relative exchange errors $|\Delta E_x|=|(E_x^{exact}-E_x^{approx})/E_x^{LDA}|$ for jellium clusters of bulk 
parameter $r_s=4$ and up to $Z=2048$.}
\label{f1b}
\end{figure}

In Fig. \ref{f1b} we have plotted the relative error
$\Delta E_x=(E_x^{exact}-E_x^{approx})/E_x^{LDA}$, computed over the volume 
($V$ defined as in Eq. (\ref{e2})), for jellium clusters of bulk 
parameter $r_s=4$ and up to $Z=2048$ electrons. 
The restriction of the intergal domain to the volume $V$ allows to remove 
the non-integrable region for the fourth-order terms (i.e. the tail of the 
density) and to consider solely the slowly-varying density region.
The LDA, MGE2, and GE4 results are also reported in the figure. 

It can be seen that MGE4 behaves similarly to GE4, which
is derived from the slowly-varying density limit
behavior. Actually, MGE4 gives even the best results
for larger clusters. For these latter systems,
MGE4 also outperforms MGE2, which is instead
more accurate for the smallest clusters.
To our knowledge, MGE4 is the only expansion which is realistic 
for atoms (that are characterized by the Thomas-Fermi scaling \cite{scaling} 
$n_\lambda(\R)=\lambda^2 n(\lambda^{1/3}\R)$) and jellium models 
(which are models for solid-state and are based on the 
uniform-electron-gas scaling \cite{scaling} 
$n_\lambda(\R)=n(\lambda^{-1/3}\R)$).

\subsection{Construction of a generalized gradient approximation functional based on MGE4}
To demonstrate the practical utility of MGE4, 
we employ it to construct a simple 
generalized gradient approximation (GGA) functional named 
the Semiclassical GGA at fourth-order (SG4).
Being based on MGE4, we expect that 
the SG4 functional performs well
for both large atoms and solid-state systems;
moreover, we will see that it is rather accurate 
also for ordinary chemistry. 

\begin{figure}
\includegraphics[width=0.9\columnwidth]{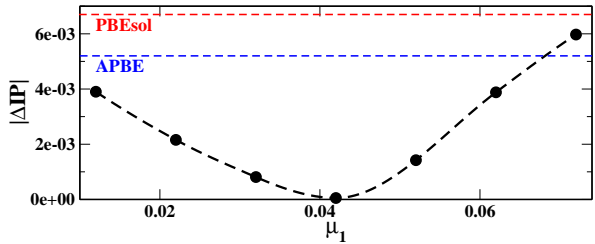}
\caption{\label{figip} Absolute error (Ha) for the exchange ionization potential in the SCA limit as a function of the parameter $\mu_1$. The errors of the PBEsol and APBE functionals are also reported for reference.} 
\end{figure}

The SG4 exchange enhancement factor takes the form
\begin{equation}
F_x^{SG4}=1+\kappa_1+\kappa_2-\frac{\kappa_1 
(1-\mu_1s^2/\kappa_1)}{1-(\mu_1s^2/\kappa_1)^5}-\frac{\kappa_2}{1+\mu_2s^2/\kappa_2},
\label{e4}
\end{equation}
where the condition
$\kappa_1+\kappa_2=0.804$ is fixed from the Lieb-Oxford bound 
\cite{lieb81}, while 
$\mu_1+\mu_2=\mu^{MGE2}=0.26$ and $\kappa_2=-\mu_2^2/\nu^{MGE4}$ 
are imposed to recover MGE2 and MGE4, respectively.
Note that in a Taylor expansion around $s=0$, the fourth term 
on the right hand side of Eq. (\ref{e4}) contributes 
only with $\mu_1s^2+\mathcal{O}(s^{10})$, whereas the MGE4
behavior is completely described by the last term. 
Such a simple splitting 
allows a better understanding of the physics behind the functional.
It remains only one free parameter 
not fixed by the previous slowly-varying density 
conditions. We fix it to $\mu_1=0.042$ by fitting to the exchange
ionization potential in the SCA limit (see Fig.
\ref{figip}).

\begin{figure}
\includegraphics[width=0.9\columnwidth]{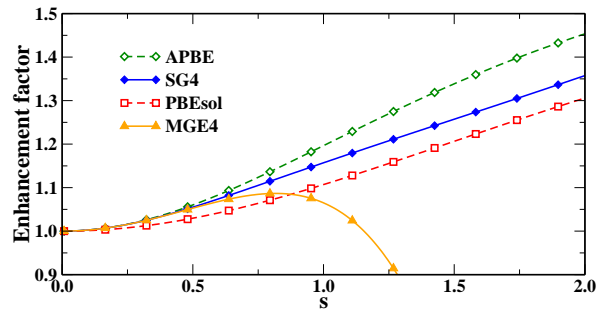}
\caption{\label{figen}Enhancement factors of different functionals as functions of the reduced gradient $s$. } 
\end{figure}

In such a way, the SG4 exchange functional is
completely constructed from the SCA model.
Its enhancement factor is reported in Fig. \ref{figen}. 
For small values of the reduced gradient $s$, it is close
to the APBE one, since in this case MGE2 and MGE4 are very similar
(they coincide in the limit of very small $s$ values).
However, unlike APBE, the SG4 functional recovers
MGE4 until $s\approx 0.6$.
For larger values of the reduced gradient the
SG4 enhancement factor is between the APBE and the PBEsol ones.

To complete the SG4 functional we need to complement it with
a correlation functional. This must describe
accurately the SCA correlation expansion
$E_c\approx 0.02072 Z\ln(Z)+B Z+\ldots$ (where 
$B=0.038$ is the best estimate for the first-order coefficient \cite{kc}),
and recover the APBE correlation in the core of
a large atom (i.e. for $r_s\rightarrow 0$ and $s\rightarrow 0$; 
note that, for exchange, SG4$\rightarrow$APBE in this limit).
Hence, we consider
the simple correlation energy per particle  
\begin{equation}
\epsilon_c^{SG4}=\epsilon_c^{LDA}+\phi^{\alpha t^3}H(r_s,\zeta,t),
\label{e5}
\end{equation}
where $t=|\nabla n|/(2 k_s\phi  n)$ is the reduced gradient for
correlation \cite{pbe}, with $k_s=(4k_F/\pi)^{1/2}$ 
being the Thomas-Fermi screening wave vector ($k_F=(3\pi^2n)^{1/3}$),
$\phi={\left[\left(1+\zeta\right)^{2/3} + \left(1-\zeta\right)^{2/3} \right]}/2$ 
is a spin-scaling  factor, $\zeta=(\UP-\DN)/n$ is 
the relative spin polarization, and $H$ is a 
localized PBE-like gradient correction 
\cite{pbe,tpssloc} where we use 
\begin{equation}
\beta=\beta_0+\sigma t (1-e^{-r_s^2})\ .
\label{e6}
\end{equation}
In order to recover the accurate LDA linear response\cite{APBE,mukappa}, we fix 
$\beta_0=3\mu^{MGE2}/\pi^2$. Moreover, we fix the parameter 
$\sigma=0.07$ fitting to jellium surface exchange-correlation 
energies \cite{js} (in analogy to PBEsol \cite{PBEsol}) and $\alpha=0.8$ 
minimizing the information entropy function described in Ref. \onlinecite{zint}. 
We recall that the spin-correction factor $\phi^{\alpha t^3}$ is 
always equal to one for spin-unpolarized systems (e.g. non magnetic
solids), being important only in the rapidly-varying spin-dependent 
density regime (e.g. small atoms) \cite{zint}.

Equations (\ref{e4}) and (\ref{e5}) define the SG4 exchange-correlation 
functional which satisfies, with no empirical parameters, 
many exact properties, including the constraints derived from the
SCA theory.

\section{Results}
In this section we present a general assessment of the
performance of the SG4 functional for 
solid-state problems, which is the main topic of this work.
For completeness, several atomic and molecular tests are 
also reported.
Finally, we consider some application examples, to show
the practical utility of MGE4 and the related SG4 functional
in condensed-matter physics.

\subsection{General assessment}
\begin{figure}
\includegraphics[width=0.9\columnwidth]{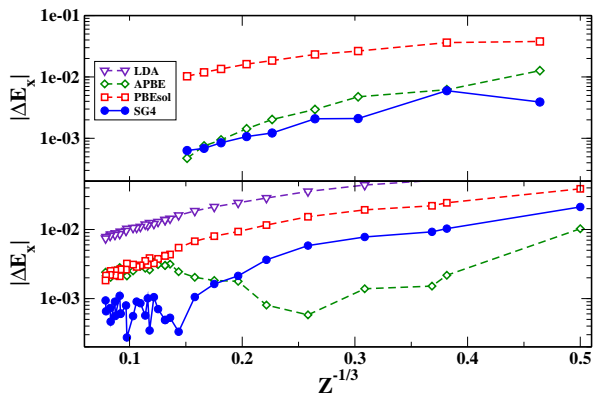}
\caption{Relative exchange errors ($\Delta E_x=|E_x^{exact}-E_x^{approx}|/E_x^{LDA}$) 
for noble-gas atoms up to $Z=290$ electrons (upper panel) and for jellium clusters (lower panel) with $r_s=4$, up to $Z=2048$.}
\label{f2}  
\end{figure}

At first, we consider a general assessment of the exchange only
SG4 functional. This will allow a more direct evaluation of the
importance of the MGE4 recovery. 
In Fig. \ref{f2} we show the relative errors 
$\Delta E_x=|E_x^{exact}-E_x^{approx}|/E_x^{LDA}$ 
for noble atoms (upper panel) and jellium clusters (lower panel),
for several exchange functionals.
The PBEsol exchange, which is not based on the SCA theory, 
is not accurate for atoms \cite{PCSB}, while APBE and SG4 
are very accurate. 
This result is not highly surprising, since both these functionals
are constructed to recover the SCA theory. Nevertheless, 
it is interesting to note that a good performance
is obtained not only for very large atoms, but also for the moderately
small ones.
In the case of jellium clusters, APBE is the best for $Z<100$, while
SG4 becomes more accurate for larger values of $Z$. 
According to the liquid-drop model theory of 
jellium spheres \cite{scaling}, this means that it describes
accurately the exchange quantum effects present in these systems.

Next, we discuss the performance of the full 
exchange-correlation SG4 functional
for some basic solid-state tests. For completeness, several molecular 
tests are also reported.
\begin{table}
\begin{center}
\caption{\label{tab1} Mean absolute errors for equilibrium lattice constants and bulk moduli of a set of 29 bulk materials and for several chemistry tests [in detail, atomization energies of main-group molecules (G2/97) and transition metal complexes (TM10AE), metal-organic interfaces (SI12), bond lengths involving H atoms (MGHBL9) and not involving H atoms (MGNHBL11), interaction energies of hydrogen bond and dipole-dipole complexes (HB6+DI6), interaction energies of dihydrogen bond complexes (DHB23)]. The best (worst) results of each line are in boldface (underlined).}
\begin{ruledtabular}
\begin{tabular}{lrrrrr}
                  & APBE & PBE & SG4 & PBEsol & WC  \\
\hline
\multicolumn{6}{c}{Lattice constants (m\AA)} \\
simple metals     & 34.1 & \bf{31.4} & 41.7 & 55.6 & \underline{58.4} \\
transition metals & \underline{63.7} & 44.9 & 23.9 & 25.6 & \bf{23.7} \\
semiconductors    & \underline{102.3} & 85.3 & \bf{21.0} & 32.7 & 31.5 \\
ionic solids      & \underline{95.6} & 76.0 & 20.8 & 20.0 & \bf{18.8}  \\
insulators        & \underline{34.0} & 27.8 & {\bf 8.0} &  8.5 & 8.0 \\
total MAE         & \underline{66.0} & 53.0 & {\bf 24.9} & 31.0 & 30.7 \\
LC20              & \underline{70.3} & 55.9 & {\bf 23.1}  &  34.0 & 31.8 \\
\multicolumn{6}{c}{bulk moduli (GPa)} \\
simple metals     & \underline{1.3} & 1.1 &  0.7 &  { \bf 0.2} & 0.54  \\
transition metals & \underline{26.3} & 21.2 & 20.0 & 20.2 & \bf{18.9} \\
semiconductors    & \underline{18.7} & 16.9 & {\bf 5.8}&  8.2 & 8.0  \\
ionic solids      & \underline{9.6} & 8.5 & 6.2 & {\bf 3.9} & 4.8 \\
insulators        & \underline{18.6} & 15.3 & {\bf 4.9} &  6.2 & 6.1 \\
total MAE         & \underline{14.8} & 12.4 & {\bf 7.9} & 8.2  & 8.0 \\
\multicolumn{6}{c}{cohesive energies (eV)} \\
simple metals     & 0.09 & \bf{0.05} &  \underline{0.21} & 0.15 & 0.13 \\
transition metals & 0.32 &\bf{0.21} & 0.39 & \underline{0.62} & 0.51 \\
semiconductors    & 0.29 & 0.13 & {\bf 0.09} & \underline{0.28} & 0.20 \\
ionic solids      & 0.19 & 0.14 & \underline{0.20} & 0.07 & \bf{0.06}  \\
insulators        & \bf{0.12} & 0.16 & 0.36 & \underline{0.57} & 0.47 \\
total MAE         & 0.21 & \bf{0.14} & 0.24 & \underline{0.33} & 0.27 \\
\multicolumn{6}{c}{Molecular tests (kcal/mol ; m\AA)} \\
G2/97 & {\bf 8.9}  & 14.8 & 15.7 & \underline{37.7} & 27.6 \\
TM10AE & {\bf 11.1} & 13.0 & 11.9 & \underline{18.3} & 15.8 \\ 
SI12 & \underline{5.9} & 3.7 & {\bf 2.6} & 3.8 & 3.3 \\   
MGHBL9 & {\bf 10.0} & 11.5 & 10.3 & \underline{14.5} & 13.9 \\
MGNHBL11  & \underline{9.2}  & 7.6  & {\bf 3.7}  & 5.2 & 6.1 \\
HB6+DI6   & {\bf 0.4}  & {\bf 0.4}  & 0.5  & \underline{1.3} & 0.9 \\ 
DHB23     & {\bf 0.8} & 1.0 & 1.1 & \underline{1.8} & 1.6 \\
\end{tabular}
\end{ruledtabular}
\end{center}
\end{table}
In Table \ref{tab1} we show the results of SG4 
calculations for the lattice constants, bulk moduli, and cohesive energies 
of a set of 29 bulk solids (see Section \ref{compdet}).
Our benchmark set for lattice constants includes, as a subset, the 
LC20 benchmark set of Refs. \onlinecite{coads,SCAN}, which is also 
reported in Table \ref{tab1}. 
The comparison is done with APBE \cite{APBE}, that is 
the other non-empirical XC functional based on the SCA theory, as well as 
with the PBEsol \cite{PBEsol}, PBE \cite{pbe}, and Wu-Cohen (WC) \cite{wc} 
functionals, which are among the most popular GGAs for solids
(another popular solid-state functional is the AM05 \cite{am05,am05_2,am05_3}
(not reported), which performs similarly to PBEsol and WC).

It can be seen that SG4 works remarkably well
for solids.
It outperforms APBE (and PBE) and is often even better than 
the state-of-the-art GGA for solids PBEsol and WC.
The comparison of the SG4 results for lattice constants
and bulk moduli with the APBE ones shows the relevance of MGE4
for solid-state systems.
We highlight that the SG4 result for the LC20 test (MAE=23.1 m\AA) also 
competes with the ones of the best meta-GGAs for solids. 
From literature, we found indeed the following MAEs for the LC20 test set:
 TPSS    = 43 m\AA \cite{coads}, 
 revTPSS = 32 m\AA \cite{coads}, 
 SCAN    = 16 m\AA \cite{SCAN}.
This is a remarkable performance of the SG4 functional for the 
equilibrium lattice constants of bulk solids, suggesting that the 
MGE4 gradient expansion can also be a useful tool for further 
meta-GGA development.

The results for the cohesive energies display a quite different trend.
Actually, this property involves a difference between results from bulk
and atomic calculations. Thus, the best results
are found for the PBE functional, which provides the
best error cancellation (note that PBE is the best neither
for solid-state nor for atoms). The SG4 functional
performs overall similarly as the APBE one,
being slightly penalized by the need to include small atoms' 
calculations. Nevertheless, SG4 definitely outperforms
PBEsol, which yields a quite poor description of all atoms.

Cohesive energy results can be rationalized even better
looking at the outcome of several molecular tests.
These tests are also useful to obtain a more comprehensive
assessment of the performance of the functionals,
even though we recall that the focus of the present paper
is on solid-state properties.
Moreover, the comparison of SG4 with PBEsol
provides a hint of the relevance of the SCA theory 
underlying the SG4 construction.

Inspection of the lowest panel of Table \ref{tab1}
shows that SG4 is quite accurate for molecular tests. 
It is comparable with PBE for atomization and non-covalent energies
(within chemical accuracy), and very accurate for geometry and 
interaction energies at interfaces.
The latter results are especially interesting, since 
these tests require a delicate balance between the description
of different density regimes \cite{mukappa}, which
is important for broad applicability at the GGA level
\cite{PBEint,mukappa,htbs}.
In particular, the molecular bond lengths in the MGNHBL11 test 
(that do not imply bonds with hydrogen atoms) are
best described by semilocal functionals with low non-locality \cite{mukappa}
(e.g. PBEsol), while the ones in the MGHBL9 test require a
large amount of non-locality (APBE works at best).
SG4 appears to be able to capture well both situations and yields a
total mean absolute error (MAE) for geometry of 6.7 m\AA{}, better than 
both APBE (9.5 m\AA) and PBE (9.4 m\AA).

\subsection{Surface and monovacancy formation energies}
\label{subb}
\begin{table}
\begin{center}
\caption{\label{tab2} 
(111) surface energies (J/m$^2$) and monovacancy formation energies (eV)
in several simple and transition metals. Mean absolute errors (MAE) are reported in the last line. Values in best agreement with experiments \cite{q2d,singhmiller09,luo12,vitos,delczeg09,ehrhart,schaefer87,deboer88} are in boldface.
}
\begin{ruledtabular} 
\begin{tabular}{lrrr}
Metal  & PBEsol & SG4 & Exp. \\
\hline
\multicolumn{4}{c}{Surface energies (J/m$^2$)} \\
Al & 0.96 & \bf{1.06} & 1.14 \\
Ca & \bf{0.52} & 0.54 & 0.50 \\
Sr & 0.40 & \bf{0.41} & 0.42 \\
Cu & 1.61 & \bf{1.67} & 1.79 \\
Pt & 1.83 & \bf{1.89} & 2.49 \\
Rh & 2.45 & \bf{2.51} & 2.70 \\
Au & 0.98 & \bf{1.01} & 1.50 \\
Pd & 1.69 & \bf{1.72} & 2.00 \\
MAE & 0.27 & {\bf 0.23} & \\
\multicolumn{4}{c}{Monovacancy energy (eV)} \\
Cu & \bf{1.25} & 1.35 & 1.28 \\
Ni & 1.73 & \bf{1.83} & 1.79  \\
Pd & 1.49 & \bf{1.59} & 1.85 \\
Ir & 1.88 & \bf{2.04} & 1.97 \\
Au & 0.65 & \bf{0.78} & 0.89 \\
Pt & 1.02 & \bf{1.15} & 1.35 \\
MAE & 0.19 & {\bf 0.13} & \\
\end{tabular}
\end{ruledtabular}
\end{center}
\end{table}
In Table \ref{tab2} we report the surface energies of 
three simple metals and five transition metals, as well as
the monovacancy formation energies in several transition metals.
These tests involve a comparison between bulk energies in 
a delocalized electronic system (metal)
and the energy of the quite localized surface/vacancy.
Thus, they may be the ideal playground for the SG4 functional
which shows a good performance for bulk, being simultaneously
quite accurate also for confined systems thanks to the
underlying SCA theory.

Indeed, SG4 performs remarkably well for both problems, yielding MAEs of
0.23 J/m$^2$ and 0.13 eV, which compare favorably with those
of PBEsol (0.27 J/m$^2$ and 0.19 eV). WC is close to PBEsol 
but slightly worse (MAEs are 0.29 J/m$^2$ and 0.22 eV);
PBE and APBE are systematically worse than PBEsol and are not reported. 
Notably, the improvement is also systematic, 
since SG4 is always closer to the experimental values
than PBEsol, with the only exception of Ca and Cu for surface and monovacacy formation
energies, respectively.

\subsection{Structure of boehmite and diaspore crystals}
\label{subc}
\begin{figure}
\includegraphics[width=0.9\columnwidth]{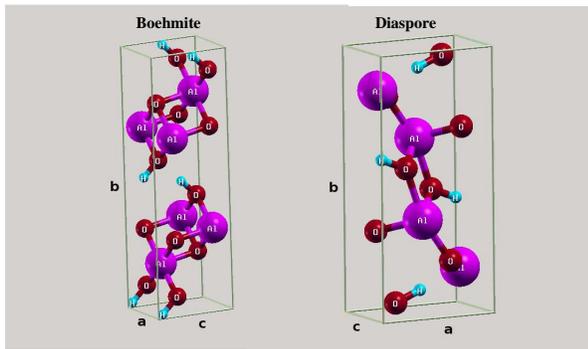}
\caption{\label{fig_boehmite} Structural parameters of boehmite and diaspore crystals.}
\end{figure}

In Table \ref{tab5} we list the structural parameters,
as defined in Fig. \ref{fig_boehmite},
computed for the boehmite and diaspore crystals.
\begin{table}
\begin{center}
\caption{\label{tab5} Lattice parameters ($a$, $b$, $c$) and various atomic distances for the boehmite and diaspore crystals. All data are in \AA. The reference data are taken from Refs. \onlinecite{hill79,corbato85}. The best result for each line is highlighted in bold.}
\begin{ruledtabular}
\begin{tabular}{lrrr}
     & PBEsol & SG4 & Exp. \\
\hline
\multicolumn{4}{c}{Boehmite} \\
a    &  \bf{2.868} &  2.863 &  2.868 \\
b    & 11.839 & \bf{11.858} & 12.234 \\
c    &	3.714 &	 \bf{3.713} &  3.692 \\
Al-O &	1.911 &  \bf{1.909} &  1.907 \\
OH   &	1.027 &  \bf{1.019} &  0.970 \\
H$\cdots$O & 1.545 & \bf{1.565} & 1.738 \\
\multicolumn{4}{c}{Diaspore} \\
a    &  4.360 &  \bf{4.362} &  4.401 \\
b    & \bf{9.411} & 9.399 & 9.425 \\
c    & \bf{2.848} &  \bf{2.842} &  2.845 \\
Al-O &  1.917 &  \bf{1.915} &  1.915 \\
OH   &  1.039 &  \bf{1.031} &  0.989 \\
H$\cdots$O & 1.524 & \bf{1.535} & 1.676 \\
\end{tabular}
\end{ruledtabular}
\end{center}
\end{table}
These systems consist of layers of aluminum hydroxides bound together by
hydrogen bonds. We recall that layered solids are becoming increasingly important
in materials science applications, thanks to their
anisotropic behavior. However, an accurate description of the equilibrium structure
of these systems requires the ability to 
describe both covalent and non-covalent bonds in the bulk
with similar accuracy. This is a quite difficult task for GGAs \cite{demichelis10}.

In general, the PBEsol functional is among the best GGAs for
boehmite and diaspore \cite{demichelis10}. 
It describes with good accuracy the covalent bonds,
but it suffers of some limitations in the description
of the non-covalent ones. The SG4 functional preserves the
good features of PBEsol and provides, in addition, small but important 
and systematic improvements for hydrogen bonds.

\subsection{Ice lattice mismatch problem}
\label{subd}
The ice lattice mismatch problem is a
popular problem in solid-state physics
\cite{feibelman08,fang13}.
It involves the calculation of the lattice constant
$a$ of ice Ih and of the lattice
constant $b$ of $\beta$-AgI, which are
used to define the lattice mismatch
\begin{equation}
f= \frac{2(b-a)}{b+a}\ .
\end{equation}
The lattice mismatch $f$ is an important quantity
in many applications, since it determines the 
growth rate of ice on a $\beta$-AgI
surface (for example, $\beta$-AgI is
used as seed crystal to produce artificial rainfall).
However, the computational determination of
the lattice mismatch is a quite hard task, since
it involves the simultaneous calculation
of the lattice constants of two 
materials with quite different electronic properties.
For this reason its calculation is 
a challenge for semilocal density functionals
\cite{feibelman08,fang13}.

In Table \ref{tabice} we report the computed 
lattice constants and the corresponding lattice mismatch,
as obtained from several functionals.
In this case we also report PBE results, since
for this problem PBE is one of the best GGAs, performing 
better than PBEsol.
\begin{table}
\begin{center}
\caption{\label{tabice}Calculated lattice constants (\AA) for ice Ih and $\beta$-AgI as well as the corresponding lattice mismatch. Experimental data are taken from Ref. \onlinecite{feibelman08}. The best result for each line is highlighted in bold.}
\begin{ruledtabular}
\begin{tabular}{lrrrr}
 & PBE & PBEsol & SG4 & Exp. \\
\hline
a (ice Ih) & {\bf 4.42} & 4.29 & 4.33 & 4.50 \\
b ($\beta$-AgI) & 4.68 & 4.56 & {\bf 4.57} & 4.59 \\
mismatch & 5.7\% & 6.1\% & {\bf 5.4\%} & 2.2\% \\
\end{tabular}
\end{ruledtabular} 
\end{center}
\end{table}
Indeed, we see that PBE yields quite good results
for the lattice constant of both materials,
showing errors below 0.1 \AA{} in both cases.
However, because the errors
for ice Ih and $\beta$-AgI lattice constants 
have opposite signs, the final lattice mismatch is
computed rather inaccurately.
On the other hand, PBEsol performs well for $\beta$-AgI,
but yields a much larger error for ice Ih. Thus,
the resulting lattice mismatch is definitely 
overestimated. 
A better balance in the performance is seen instead
in the SG4 case. This functional is in fact
the best for $\beta$-AgI and between PBE and PBEsol
for ice Ih. Hence, it finally yields a lattice
mismatch of 5.4\%. This value is still
too large with respect to the experimental one
(2.2\%), but improves with respect to the other
GGAs.


\section{Summary and concluding remarks}
In this paper we have used the SCA theory to introduce 
a modified fourth-order gradient expansion (MGE4), 
which is accurate for atoms and solid-state systems.
The MGE4 includes and extends the well-known MGE2.
The extension over MGE2 is obtained by using, in the
gradient expansion construction, an additional constraint, 
beyond the energy one, which provides an improved
description of slowly-varying regions of different materials.
To implement the new constraint, an integration over
slowly-varying density regions was introduced in order to 
reduce computational noise and avoid the divergence of fourth order
terms in rapidly-varying density regions.  

To exploit the good features of MGE4 we have used it
as a base to construct a simple GGA functional, named
SG4. This functional is free of parameters fitted on 
real systems and satisfies several exact properties, including
those relevant for the SCA model, the Lieb-Oxford bound,
the LDA linear response behavior, and the rapidly-varying as well 
as the high-density limits of correlation.
The SG4 functional performs well for a broad range of 
problems in solid-state physics, still preserving, thanks to
the SCA underlying theory, a reasonable performance also
for molecular tests.
Due to this eclectic character, the SG4 functional
is particularly promising for problems involving multiple
electronic structure features, such as surface energies, 
vacancies, non-covalent interactions in bulk solids,
and interfaces.

The results of the present work highlight, through
the power of MGE4 for different problems, the
importance of the underlying SCA model as a 
reference system in DFT
also for solid-state systems. 
This was the primary goal of the present work, since
the relevance of the SCA model system was often overlooked in 
the literature
and the utility of this model has been often considered to be 
limited to the atomic and molecular framework.

To conclude we recall that gradient expansions are basic tools 
for the contruction of non-empirical DFT functionals, even beyond the
semilocal XC level 
\cite{TPSS,Becke,ernzerhof98,final1,final2,final3,final4}.
Thus, in the future, further studies may focus on the use
of MGE4 to construct and optimize highly accurate 
functionals of different ranks,
beyond the simple SG4 that we have presented here mainly
to illustrate the practical utility of the SCA theory.

\section{Computational details}
\label{compdet}
All atomic calculations used to derive the MGE4 expansion
have been performed using the Engel code \cite{engel}
with the exact exchange functional. Further tests of the MGE4 
on jellium clusters have been carried out employing 
accurate LDA Kohn-Sham densities \cite{tao08}.

The SG4 functional has been tested on different datasets, including
\begin{itemize}
\item Atomization and interaction energies: atomization energies of
small molecules (G2/97 \cite{g2_1,g2_2}), atomization energies of small transition
metal complexes (TM10AE \cite{furche06,zint,zvint}), 
small gold-organic interfaces (SI12 \cite{hpbeint}).
\item Structural properties: bond lengths involving H atoms (MGHBL9 \cite{mgbl19}),
bond lengths not involving H atoms (MGNHBL11 \cite{mgbl19}) bonds.
\item Non-covalent interactions: interaction energies of hydrogen-bond
and dipole-dipole complexes (HB6+DI6 \cite{zhao05}) as well as
of dihydrogen bond complexes (DHB23 \cite{dihydro}).
\item Solid-state tests: Equilibrium lattice constants and bulk moduli of
29 solids, including Al, Ca, K, Li, Na, Sr, Ba (simple metals); Ag, Cu,
Pd, Rh, V, Pt, Ni (transition metals); LiCl, LiF, MgO, NaCl, NaF
(ionic solids); AlN, BN, BP, C (insulators); GaAs, GaP, GaN, Si, SiC,
Ge (semiconductors). Reference data to construct this set were taken
from Refs. \onlinecite{coads,harl10,schimka11,csonka09,janthon14}.
\end{itemize}
All calculations for molecular systems have been performed with the
TURBOMOLE program package \cite{turbomole,turbo_rev}, using a def2-TZVPP basis set
\cite{basis1,basis2}. 
Calculations concerning solid-state tests have been performed with the
VASP program \cite{vasp}, using PBE-PAW pseudopotentials. 
We remark that the use of the same pseudopotential
for all the functionals may lead to inaccuracies in the final results.
Nevertheless,
the use of PAW core potentials ensures good transferability for multiple functionals
\cite{am05_3,paier05}, 
since the core-valence interaction is recalculated for each functional. Indeed,
test calculations employing different variants of the PAW potentials
(GGA-PAW)
have shown that the estimated convergence level of our calculations is
about 1 m\AA{} for lattice constants, 0.5 GPa for bulk moduli, 
and 0.01 eV for cohesive energies. All Brillouin zone integrations
were performed on $\Gamma$-centered symmetry-reduced Monkhorst-Pack
$k$-point meshes, using the tetrahedron method with Bl\"ochl corrections. For
all the calculations a $24\times 24 \times 24$ $k$-mesh 
grid was applied and the plane-wave
cutoff was chosen to be 30\% larger than maximum cutoff defined for the
pseudopotential of each considered atom. The bulk modulus was obtained
using the Murnaghan equation of state.
The cohesive energy, defined as the energy per atom needed to atomize the
crystal, is calculated for each functional from the energies of the crystal at
its equilibrium volume and the spin-polarized symmetry-broken solutions
of the constituent atoms. To generate symmetry breaking solutions, atoms
were placed in a large orthorhombic box with dimensions $13\times14\times15$ 
\AA$^3$.

Calculations for the examples reported in subsections \ref{subb}, \ref{subc},
and \ref{subd} have been
performed using the same computational setups as in Refs. 
\onlinecite{luo12,delczeg09,demichelis10,feibelman08}.

\textbf{Acknowledgments}. We thank Prof. K. Burke for useful discussions and
TURBOMOLE GmbH for providing the TURBOMOLE program.
E. Fabiano acknowledges a partial funding of this work 
from a CentraleSup\'elec visiting professorship.


\begin{thebibliography}{[Vo]}
%
\bibitem{hk} P. Hohenberg and W. Kohn, Phys. Rev. \textbf{136}, B864 (1964).
%
\bibitem{KS} W. Kohn and L. J. Sham, Phys. Rev. $\mathbf{140}$, A1133 (1965).
%
\bibitem{dobson_vignale_book} J. F. Dobson, G. Vignale, and M. P. Das, \textit{Electronic Density Functional Theory}, Springer (1998).
%
\bibitem{scusrev} G. E. Scuseria and V. N. Staroverov, Progress in the development of exchange-correlation functionals, Chapter 24 in: Theory and Applications of Computational Chemistry: The First 40 Years (A Volume of Technical and Historical Perspectives), edited by C. E. Dykstra, G. Frenking, K. S. Kim, and G. E. Scuseria, Elsevier, Amsterdam (2005).
%
\bibitem{rappoport09} D. Rappoport, N. R. M. Crawford, F. Furche, and K. Burke, Approximate Density Functionals: Which Should I Choose?, in Encyclopedia of Inorganic Chemistry,ohn Wiley \& Sons (2009), DOI: 10.1002/0470862106.ia615.
%
\bibitem{civalleri} B. Civalleri, D. Presti, R. Dovesi, and A. Savin, On choosing the best density functional approximation, in Chemical Modelling : Applications and Theory Volume 9, The Royal Society of Chemistry (2012).
\bibitem{PBEsol} J. P. Perdew, A. Ruzsinszky, G. I. Csonka, O. A. Vydrov, G. E. Scuseria, L. A. Constantin, X. Zhou, and K.  Burke, Phys. Rev. Lett. $\mathbf{100}$, 136406 (2008); ibid. $\mathbf{102}$, 039902 (E); ibid. $\mathbf{101}$, 239702 (2008).
%
\bibitem{sogga} Y. Zhao and D. G. Truhlar, J. Chem. Phys. \textbf{128}, 184109 (2008).
%
\bibitem{sogga11} R. Peverati, Y. Zhao and D. G. Truhlar, J. Phys. Chem. Lett. \textbf{2}, 1991 (2011)
%
\bibitem{PBEint} E. Fabiano, L. A. Constantin, and F. Della Sala, Phys. Rev. B \textbf{82}, 113104 (2010).
%
\bibitem{vt84} A. Vela, J. C. Pacheco-Kato, J. L. G\'azquez, J. M. del Campo, and S. B. Trickey, J. Chem. Phys.\textbf{136}, 144115 (2012).
%
\bibitem{vela09} A. Vela, V. Medel, and S. B. Trickey, J. Chem. Phys. \textbf{130}, 244103 (2009).
\bibitem{pedroza09} L. S. Pedroza, A. J. R. da Silva, and K. Capelle, Phys. Rev. B \textbf{79}, 201106(R) (2009).
%
\bibitem{RGE2} A. Ruzsinszky, G. I. Csonka, and G. E. Scuseria, J. Chem. Theory Comput. $\mathbf{5}$, 763 (2009).

\bibitem{inttca} E. Fabiano, L. A. Constantin, A. Terentjevs, F. Della Sala, P. Cortona, Theor. Chem. Acc. \textbf{134}, 139 (2015).
%

\bibitem{TPSS} J. Tao, J. P. Perdew, V. N. Staroverov, and G. E. Scuseria, Phys. Rev. Lett. $\mathbf{91}$, 146401 (2003).
%
\bibitem{revTPSS} J.P. Perdew, A. Ruzsinszky, G. I. Csonka, L. A. Constantin, and J. Sun, Phys. Rev. Lett.  $\mathbf{103}$, 026403 (2009); (Erratum) Phys. Rev. Lett. $\mathbf{106}$, 179902(E) (2011).
%
\bibitem{BLOC} L.A. Constantin, E. Fabiano, F. Della Sala, J. Chem. Theory Comput.  $\mathbf{9}$, 2256 (2013).
%
\bibitem{MGGAMS2} J. Sun, R. Haunschild, B. Xiao, I.W. Bulik, G. E. Scuseria, and J. P. Perdew, J. Chem. Phys. $\mathbf{138}$, 044113 (2013). 
%
\bibitem{metavt84} J. M. del Campo, J. L. G{\'a}zquez, S. B. Trickey , A. Vela, Chem. Phys. Lett. 543, 179 (2012).
%
\bibitem{SCAN} J. Sun, A. Ruzsinszky, and J.P. Perdew, Phys. Rev. Lett. $\mathbf{115}$, 036402 (2015).
%
\bibitem{M11L} R. Peverati and D.G. Truhlar, J. Phys. Chem. Lett.  \textbf{3}, 117 (2012)
%
%
%
%
%
%
%
%

\bibitem{thomas26} L. H. Thomas, Proc. Cambridge Phil. Soc. $\mathbf{23}$, 542 (1926).
%
\bibitem{fermi27} E. Fermi, Rend. Accad. Naz. Lincei $\mathbf{6}$, 602 (1927).
%
\bibitem{scott52} J. M. C. Scott, Philos. Mag. $\mathbf{43}$, 859 (1952).
%
\bibitem{schwinger80} J. Schwinger, Phys. Rev. A $\mathbf{22}$, 1827 (1980).
%
\bibitem{schwinger81} J. Schwinger, Phys. Rev. A $\mathbf{24}$, 2353 (1981).
%
\bibitem{englert84} B.-G. Englert and J. Schwinger, Phys. Rev. A $\mathbf{29}$, 2339 (1984).
%
\bibitem{englert85} B.-G. Englert and J. Schwinger, Phys. Rev. A $\mathbf{32}$, 26 (1985).
%
\bibitem{EB09} P. Elliott and K. Burke, Can. J. Chem. $\mathbf{87}$, 1485 (2009).
%
\bibitem{ELCB} P. Elliott, D. Lee, A. Cangi, and K. Burke, Phys. Rev. Lett. $\mathbf{100}$, 256406 (2008).
%
\bibitem{lee09} D. Lee, L. A. Constantin, J. P. Perdew, and K. Burke, J. Chem. Phys. $\mathbf{130}$, 034107 (2009).
%
\bibitem{PCSB} J. P. Perdew, L. A. Constantin, E. Sagvolden, and K. Burke, Phys. Rev. Lett. $\mathbf{97}$, 223002 (2006).
%
\bibitem{APBE} L. A. Constantin, E. Fabiano, S. Laricchia, and F. Della Sala, Phys. Rev. Lett. $\mathbf{106}$, 186406 (2011).
%
\bibitem{hapbe} E. Fabiano, L. A. Constantin, P. Cortona, and F. Della Sala, J. Chem. Theory Comput. \textbf{11}, 122 (2015).
%
\bibitem{ioni} L. A. Constantin, J. C. Snyder, J. P. Perdew, and K. Burke, J. Chem. Phys. $\mathbf{133}$, 241103 (2010).
%
%
\bibitem{lieb81} E. H. Lieb and S. Oxford, Int. J. Quantum Chem. \textbf{19}, 427 (1981).
%
\bibitem{sLL} L. A. Constantin, A. Terentjevs, F. Della Sala, and E. Fabiano, Phys. Rev. B $\mathbf{91}$, 041120(R) (2015).
\bibitem{madsen} G. K. H. Madsen, Phys. Rev. B {\bf 75}, 195108 (2007)  

\bibitem{wc} Z. Wu and R. E. Cohen, Phys. Rev. B \textbf{73}, 235116 (2006). 

\bibitem{wcomm} Y. Zhao and D.G. Truhlar  \textbf{78}, 197101 (2008)
\bibitem{alpha} F. Della Sala, E. Fabiano, L. A. Constantin, Phys. Rev. B \textbf{91}, 035126 (2015).
%
\bibitem{pb} J. P. Perdew, V.N. Staroverov, J. Tao, and G. E. Scuseria, Phys. Rev. A $\mathbf{78}$, 052513 (2008).
%
\bibitem{Becke} A. D. Becke, J. Chem. Phys. $\mathbf{104}$, 1040 (1996). 
%
\bibitem{ernzerhof98} M. Ernzerhof and J. P. Perdew, J. Chem. Phys. $\mathbf{109}$, 3313 (1998).
%
\bibitem{hole} L. A. Constantin, E. Fabiano, and F. Della Sala, Phys. Rev. B \textbf{88}, 125112 (2013).
%
\bibitem{scaling} E. Fabiano and L. A. Constantin, Phys. Rev. A $\mathbf{87}$, 012511 (2013).
%
\bibitem{kc} K. Burke, A. Cancio, T. Gould, and S. Pittalis, arXiv:1409.4834 [cond-mat.mtrl-sci].
%
\bibitem{pbe} J. P. Perdew, K. Burke, and M. Ernzerhof, Phys. Rev. Lett. $\mathbf{77}$, 3865 (1996).
%
\bibitem{tpssloc} L. A. Constantin, E. Fabiano, and F. Della Sala, Phys. Rev. B $\mathbf{86}$, 035130 (2012).
%
\bibitem{mukappa} E. Fabiano, L. A. Constantin, and F. Della Sala, J. Chem. Theory Comput. \textbf{7}, 3548 (2011). 
%
\bibitem{js} L. A. Constantin, L. Chiodo, E. Fabiano, I. Bodrenko, F. Della Sala, Phys. Rev. B \textbf{84}, 045126 (2011).
%
\bibitem{zint} L. A. Constantin, E. Fabiano, and F. Della Sala, Phys. Rev. B $\mathbf{84}$, 233103 (2011).
%
\bibitem{am05} R. Armiento and A. E. Mattsson, Phys. Rev. B \textbf{72}, 085108 (2005).
%
\bibitem{am05_2} A. E. Mattsson and R. Armiento, Phys. Rev. B \textbf{79}, 155101 (2009).
%
\bibitem{am05_3} A. E. Mattsson, R. Armiento, J. Paier, G. Kresse, J. M. Wills, and T. R. Mattsson, J. Chem. Phys. \textbf{128}, 084714 (2008).
%
\bibitem{paier05} J. Paier, R. Hirschl, M. Marsman, and G. Kresse, J. Chem. Phys. \textbf{122}, 234102 (2005).
%
\bibitem{coads} J. Sun, M. Marsman, G. I. Csonka, A. Ruzsinszky, P. Hao, Y.-S. Kim, G. Kresse, and J. P. Perdew, Phys. Rev. B \textbf{84}, 035117 (2011).
%
%
\bibitem{htbs} P. Haas, F. Tran, P. Blaha, and K. Schwarz, Phys Rev. B \textbf{83}, 205117 (2011).
%
\bibitem{q2d} L. Chiodo, L. A. Constantin, E. Fabiano, and F. Della Sala, Phys. Rev. Lett. \textbf{108}, 126402 (2012).
%
\bibitem{singhmiller09} N. E. Singh-Miller and N. Marzari, Phys. Rev. B \textbf{80}, 235407 (2009)
%
\bibitem{luo12} S. Luo, Y. Zhao, and D. G. Truhlar, J. Phys. Chem. Lett. 3, 2975 (2012).
%
\bibitem{vitos} L. Vitos, A. V. Ruban, H. L. Skriver, and J. Kollar, Surf. Sci. \textbf{411}, 186 (1998).
%
\bibitem{delczeg09} L. Delczeg, E. K. Delczeg-Czirjak, B. Johansson, and L. Vitos, Phys. Rev. B \textbf{80}, 205121 (2009).
%
\bibitem{ehrhart} P. Ehrhart, P. Jung, H. Schultz, and H. Ullmaier, \textit{Atomic Defects in Metals}, Landolt-B\" ornstein, New Series, Group III vol 25, Springer (1991).; 
%
\bibitem{schaefer87} H.-E. Schaefer, Phys. Stat. Sol. (a) \textbf{102}, 47 (1987).
%
\bibitem{deboer88} F. R. de Boer, R. Boom, W. C. M. Mattens, A. R. Miedema, and A. K. Niessen, \textit{Cohesion in Metals}, Vol. 1, North-Holland, Amsterdam (1988).
%
\bibitem{hill79} R. Hill, J. Phys. Chem. Miner. \textbf{5}, 179 (1979).
%
\bibitem{corbato85} C. E. Corbato, R. T. Tettenhorst, and G. G. Christoph, Clays Clay Miner. \textbf{33}, 71 (1985).
%
\bibitem{demichelis10} R. Demichelis, B. Civalleri, P. D'Arco, and R. Dovesi, Int. J. Quant. Chem. \textbf{110}, 2260 (2010).
%
\bibitem{feibelman08} P. J. Feibelman, Phys. Chem. Chem. Phys. \textbf{10}, 4688 (2008).
%
\bibitem{fang13} Y. Fang, B. Xiao, J. Tao, J. Sun, and J. P. Perdew, Phys. Rev. B \textbf{87}, 214101 (2013).
%
\bibitem{final1} M. M. Odashima and K. Capelle, Phys. Rev. A \textbf{79}, 062515 (2009). 
%
\bibitem{final2} R. Hanunschild, M. M. Odashima, G. E. Scuseria, J. P. Perdew, and K. Capelle, J. Chem. Phys. \textbf{136}, 184102 (2012). 
%
\bibitem{final3} G. Vignale and W. Kohn, Phys. Rev. Lett. \textbf{77}, 2037 (1996).
%
\bibitem{final4} J. F. Dobson and B. P. Dinte, Phys. Rev. Lett. \textbf{76}, 1780 (1996).
%
\bibitem{engel} E. Engel, in A Primer in Density Functional Theory, Eds. C. Fiolhais, F. Nogueira, M. A. L. Marques, Springer Berlin, pp 56–122 (2003).
%
\bibitem{tao08} J. Tao, J. P. Perdew, L. M. Almeida, C. Fiolhais, and S. K\"ummel, Phys. Rev. B \textbf{77}, 245107 (2008).
%
\bibitem{g2_1} L. A. Curtiss, K. Raghavachari, P. C. Redfern, and J. A. Pople, J. Chem. Phys. \textbf{106}, 1063 (1997).
%
\bibitem{g2_2} R. Haunschild and W. Klopper, J. Chem. Phys. \textbf{136}, 164102 (2012).
%
\bibitem{furche06} F. Furche and J. P. Perdew, J. Chem. Phys. \textbf{124}, 044103 (2006).
%
%
\bibitem{zvint}L. A. Constantin, E. Fabiano, and F. Della Sala, J. Chem. Phys. \textbf{137}, 194105 (2012).
%
\bibitem{hpbeint} E. Fabiano, L. A. Constantin, and F. Della Sala, {Int. J. Quantum Chem.} \textbf{13}, 6670 (2012)
%
\bibitem{mgbl19} Y. Zhao and D. G. Truhlar, J. Chem. Phys. \textbf{125}, 194101 (2006).
%
\bibitem{zhao05} Y. Zhao and D. G. Truhlar, J. Chem. Theory Comput. \textbf{1}, 415 (2005).
%
\bibitem{dihydro} E. Fabiano, L. A. Constantin, and F. Della Sala, J. Chem. Theory Comput. \textbf{10}, 3151 (2014).
%
\bibitem{harl10}J. Harl, L. Schimka, and G. Kresse, Phys. Rev. B \textbf{81}, 115126 (2010).
%
\bibitem{schimka11} L. Schimka, J. Harl, and G. Kresse, J. Chem. Phys. \textbf{134}, 024116 (2011).
%
\bibitem{csonka09} G. I. Csonka, J. P. Perdew, A. Ruzsinszky, P. H. T. Philipsen, S. Leb\`egue, J. Paier, O. A. Vydrov, and J. G. \'Angy\'an, Phys. Rev. B \textbf{79}, 55107 (2009).
%
\bibitem{janthon14} P. Janthon, S. Luo, S. M. Kozlov, F. Vi\~nes, J. Limtrakul, D. G. Truhlar, and F. Illas, J. Chem. Theory Comput. \textbf{10}, 38323839 (2014).
%
\bibitem{turbomole} TURBOMOLE, V6.3; TURBOMOLE GmbH: Karlsruhe, Germany, 2011. Available from http://www.turbomole.com (accessed May 2015).
%
\bibitem{turbo_rev} F. Furche, R. Ahlrichs, C. H\"attig, W. Klopper, M. Sierka, and F. Weigend, WIREs Comput. Mol. Sci. \textbf{4}, 91 (2014).
%
\bibitem{basis1} F. Weigend, F. Furche, and R. Ahlrichs, J. Chem. Phys. \textbf{119}, 1275312763
(2003).
%
\bibitem{basis2} F. Weigend and R. Ahlrichs, Phys. Chem. Chem. Phys. \textbf{7}, 3297 (2005).
%
\bibitem{vasp} G. Kresse and J. Furthmuller, Phys. Rev. B \textbf{54}, 11169 (1996).
%

\end{thebibliography}
\end{document}